\documentclass[proceedings,opts]{JHEP}
\usepackage{epsfig}
\conference{Heavy Flavors 8, Southampton, UK, 1999}
\title{NRQCD: A Critical Review}
\author{ Ira Rothstein\\ Dept. of Physics\\
Carnegie Mellon University Pittsburgh Pa.\\
15213 USA}

\abstract{In this talk I  review some recent applications of NRQCD.
I first discuss the unquenched NRQCD lattice extractions of the
strong coupling constant, paying  particular attention to the recent advances
in
reducing systematic errors.
I then discuss the progress made in testing  the
NRQCD/factorization
formalism for onium production. In particular, I gather the evidence,
or lack thereof, 
for the universality of the production matrix elements. I also discuss
the interesting question of polarized production, which is a crucial
test of the formalism. I address 
the viability of this theory for  the $J/\psi$ system as well as what
needs to be done before we can reach any definitive conclusions.
}
\begin{document}
The study of heavy quarks is interesting from a phenomenological
as well as formal standpoint. Indeed, the utility of the experiments designed
to study the CKM sector of the standard model is bounded by our
theoretical  
understanding of states containing heavy quarks. 
For theorists this
presents a great challenge given that, in general, it is quite
difficult to make predictions for strongly interacting particles
from first principles. However, we do have one powerful tool
at our disposal which saves us from despondency, namely effective
field theories. These theories represent limits of QCD which 
make approximate symmetries manifest,
thus greatly enhancing our predictive power. 
Furthermore, since the effective theory reproduces QCD in a well
defined limit, it is 
possible to calculate corrections in a systematic fashion.

The effective theory I have been charged to review is non-relativistic
QCD. As the moniker implies, this is a theory which approximates
full QCD 
when applied to a bound state containing more than one heavy
quark. The quarks
 necessarily have small velocities in the 
limit $m_Q \gg \Lambda_{QCD}$, due to asymptotic freedom. 
The theory is written down as a simultaneous
expansion in $\alpha_S(2m_Q)$ and the relative velocity in the
center of mass frame of the quarks, $v$.
Indeed, if we wish, we may calculate corrections to arbitrary
order in powers of $v$ and $\alpha_s$, at the price
of introducing unknown, yet universal parameters. Thus, the relevant 
question is  not  ``is
NRQCD  correct?'', but rather, ``how well do
the $v$ and $\alpha_s$ expansions converge for the system of interest?'' 

In this review I will discuss recent advances in the application
of NRQCD.  In particular, I will concentrate
on applications for which we have new data. Thus, due to space
limitations, I will not discuss
the interesting topic of inclusive onia decays, nor will I discuss
recent theoretical progress on threshold top quark production\cite{beneke}.
Instead, I will focus on the utilization of NRQCD to measure
the strong coupling, and the predictions for onia production.

\section{The Effective Field Theory}

The effective Lagrangian is constructed \cite{BBL} by ensuring that at the
matching scale, $m_Q$, the effective
field theory reproduces on shell Greens functions of QCD to some
fixed order in $\alpha(2m_Q)$ and $v$.
The operators in the effective field theory are then classified 
by their ``effective $v$ dimension''\cite{scale}. That is, 
to each operator we
may assign a power of $v$ which is determined by some velocity
power counting rules. For our purposes, we will be interested
in the fermion anti-fermion sector of the theory where
the lowest order Lagrangian is given
by 
\begin{eqnarray}
\label{Leff}
& &  L_0 = \int dtd^3{\bf x} \Psi^\dagger(iD_0+{{\bf
\vec{D}}^2\over{2m}})\Psi (t,\vec{{\bf x}}) +\nonumber \\
& &\int d^3{\bf y} 
\Psi^\dagger T^a \Psi(t,\vec{{\bf y}}){4 \pi \alpha_s C_V \over{\mid
\vec{\bf{x}}-\vec{\bf{y}}\mid^2}}
\chi^\dagger{\bar T}^a\chi(t,\vec{{\bf x}}) \nonumber \\
& & +\sc{L}_{g}+\sc{L}_{q} + \Psi \leftrightarrow \chi,
\end{eqnarray}
where $\psi$ and $\chi$ are the fermion and anti-fermion two spinors
respectively, $\sc{L}_{g}$ and $\sc{L}_{q}$ are the unscathed dimension
four operators of the QCD Lagrangian for the gluons and light
quarks respectively. $C_V$ is unity at tree level but is corrected
at higher order. The coefficients of the other operators in the
heavy quark sector are fixed by reparameterization invariance \cite{LM}.
Terms of higher order in $v$ are easily calculated by expanding
the full theory on-shell matrix elements in the $v$.
In the one fermion sector this leads to the usual terms encountered
in non-relativistic quantum mechanics, such as the magnetic moment
and Darwin interactions. 
There is one technical detail which I should point out at this
point.  When matching beyond leading order in the strong coupling,  
the above Lagrangian must be modified if one wishes to preserve
manifest power counting. In particular, if one wishes to
regulates integrals using dimensional regularization, then
the non-potential interactions must be dipole expanded\cite{M,GRI}.
That is to say, we must make the replacement
\begin{equation}
A_\mu (t,\vec{{\bf x}}) \rightarrow A_\mu (t,0)+\vec{\bf x}\cdot
\vec{\partial}  A_\mu (t,0)+.~.~.
\end{equation}
This dipole expansion incorporates the fact that the typical
gluon momentum is of order $mv^2$ whereas the size of the
state is order $mv$. On a calculational level this leads 
to gluonic interactions which do not  transfer three momentum,
which in turn ensures manifest power counting. 
The situation is actually slightly more complicated that this, because
there are also gluons which have typical momentum of order $mv$\cite{BS}.
This leads to further modifications of the Lagrangian, which I will
not have space to discuss in this review \cite{greise,lmr}.

\section{Lattice Extractions of $\alpha_s$} 

As far a lattice calculations are concerned, the motivation for
 effective field theories
is different  than the one discussed in the
introduction. In particular, by removing the heavy quark mass from
the theory it is possible to work with lattice spacings $a$, which
are larger than the Compton wavelength of the heavy quark $(ma)>1$. 
This alleviates the problem  of the quark literally falling
through the cracks. Indeed, the 
use of coarser lattices allows for  precision
measurements which would otherwise be presently unmanageable.

Perhaps the most interesting use of lattice NRQCD is for the purpose
of extracting the value of the strong coupling constant $\alpha_s$.
This low energy extraction is interesting, not only because it
allows us to compare to results obtained from perturbative QCD,
but also because any disparity between the low energy and  high energy 
extractions, after renormalization group running, could imply
the existence of new physics. Thus, a precise
low energy extraction is of great scientific interest. 
Below we will discuss the results of the two collaborations, ``NRQCD''
\cite{NRQCD}
and ``SESAM''\cite{SESAM}, which used NRQCD to treat the heavy quarks.
These are the first lattice extractions of $\alpha_s$ performed
in the unquenched approximation. While the NRQCD collaborations
results
have been around for several years now, the important complimentary 
results of  SESAM collaboration
are more recent. The two collaborations use
different
techniques so that their agreement would provide strong corroboration
of the results for this important measurement. 
   
The strategy behind the extraction is to first 
define a short distance 
coupling constant $\alpha_s(a)$ via some lattice observable. 
Both collaborations  utilized the so-called plaquette coupling
\cite{alphap},
which is related to small Wilson loops via
\begin{eqnarray}
\label{alp}
&&-\log W_{1,1}={4\pi \over{3}}\alpha_p(3.41/a)+\nonumber \\
& &\left(1-\alpha_p(1.1879+
(0.0249,0.070)  n_f) \right).
\end{eqnarray} 
The first number in parenthesis is for Wilson fermion and the
second is for staggered fermions, and  
\begin{eqnarray}
& & W_{m,n}=\nonumber \\
& &
{1\over{3}}\langle Re Tr\left(P \exp\left[-ig_s \oint_{m,n}
A\cdot dx \right]\right)\rangle. 
\end{eqnarray}
$W_{m,n}$ is the Wilson around a rectangular path of
size $(ma)\times(na)$.
This observable has the further advantage that its leading
non-perturbative correction,
\begin{eqnarray}
\delta W_{m,n}={-\pi a^4 (mn)^2 \over{36}}\langle \alpha_s G^2\rangle.
\end{eqnarray}
 is believed to be small due the anomalously small
value of the gluon condensate  $\langle \alpha_s G^2\rangle\simeq 
0.05~GeV^4$. Note there are no higher order perturbative correction to
the relation \ref{alp}, 
as a matter of definition. This just shuffles higher order corrections
into the relation between $\alpha_{{\overline {MS}}}$ and $\alpha_p$.
The scale $3.41/a$ is the BLM scale \cite{BLM}
determined by calculating the
$n_f$ dependent piece of the two loop perturbative expansion of
the Wilson loop. 
The lattice spacing is then determined by measuring the
 the spin independent quarkonia splittings.
These splittings are chosen
 since they are much
less sensitive to tuning of the bare heavy quark mass.
Once the value of $\alpha_p(3.41/a)$ has been extracted, it
can be related to $\alpha_{\overline {MS}}$ via
\begin{equation} 
\label{relate}
\alpha_{\overline {MS}}(Q)=\alpha_P(e^{5/6}Q)\times
\left[ 1+{2 \over{\pi}}\alpha_P+C_2 \alpha_P^2 \right]
\end{equation}
The scale $e^{5/6}Q$ is chosen to absorb the BLM piece of the
first order coefficient. The coefficient $C_2$ is only known
in the pure gauge theory\cite{LW}, in which it takes the value $C_2=0.96$.
  Using this relation its possible to   determine a value
of $\alpha^{(5)}_{\overline MS}(M_Z)$ by running with the three loop
beta function and taking into account the relevant quark thresholds.

There are several sources of errors, and the SESAM and NRQCD
collaborations incur them in different fashions, as I shall now
discuss.  As usual with lattice calculations there is the issue of
cut-off artifacts.  The NRQCD collaboration removed all $O(a^2)$ error
from the heavy quark action by using an improved action. They did not
improve the gluonic action to this order, but made a perturbative
estimate of these corrections. They then checked that their spin
splittings are independent of the lattice spacing (in the quenched
approximation) by calculating with several different spacings in the
range, $0.05-0.15~fm$, and found no variations. The use of staggered
fermions also introduces errors on the order of $O(a^2)$.  They
estimated the net effect of lattice artifact errors to be at the level
of $0.2\%$.  The SESAM collaboration used dynamical Wilson fermions
which incur lattice spacing errors at linear order in $a$.  They
worked at fixed lattice spacing and were therefore unable to perform a
scaling analysis. They quote a larger error, due to discretization,
than did the NRQCD collaboration, namely $\%5$.  This larger error can
be attributed to the difference between the light fermion actions.

In addition, there are corrections due to truncating the
Lagrangian at some fixed order in $v$. Both collaborations include
$O(v^2)$ corrections to the effective action. $O(v^4)$ corrections
should be negligibly small in the $\Upsilon$ system since the
$O(v^2)$ corrections were found to shift the mass splitting
by $10\%$. There is also an error incurred by truncation the
perturbative expansion of the Wilson coefficients for these
higher dimension operators. The coefficients of these operators
are tree level ``tadpole improved''. An estimate for this
error was made by the SESAM collaboration by varying the tadpole
improvement prescription. They found that this error is of
the same order as the relativistic errors, and estimated these
combined errors to be at $1\% $ level. Whereas the NRQCD collaboration
estimated the relativistic errors to be at the $0.2\%$ and the
errors due to tadpole prescription to be $0.5\%$.

The use of dynamical light quarks was a crucial step in making
these lattice predictions trustworthy. 
However, it is difficult, at
this time, to perform calculations with physical masses for the light
quarks. Therefore, the calculations are performed with unphysical
light quark masses, and the results are extrapolated to the physical
value. In \cite{GRII}, it was shown that the level splittings should
grow linearly with the light quark mass. The SESAM collaboration 
             calculated the splittings for several different light
quark masses, between the  radial as well
as orbital excitations. The results for the $1S-1P$ splitting are
shown in figure 1 and seem to fit a linear relationship quite well.
 The result gives us confidence that the
extrapolation to physical light quark masses is being performed 
correctly. Both collaborations estimate the error due to use
of unphysical sea quark masses to be at the $1\%$ level. 
\EPSFIGURE[h]{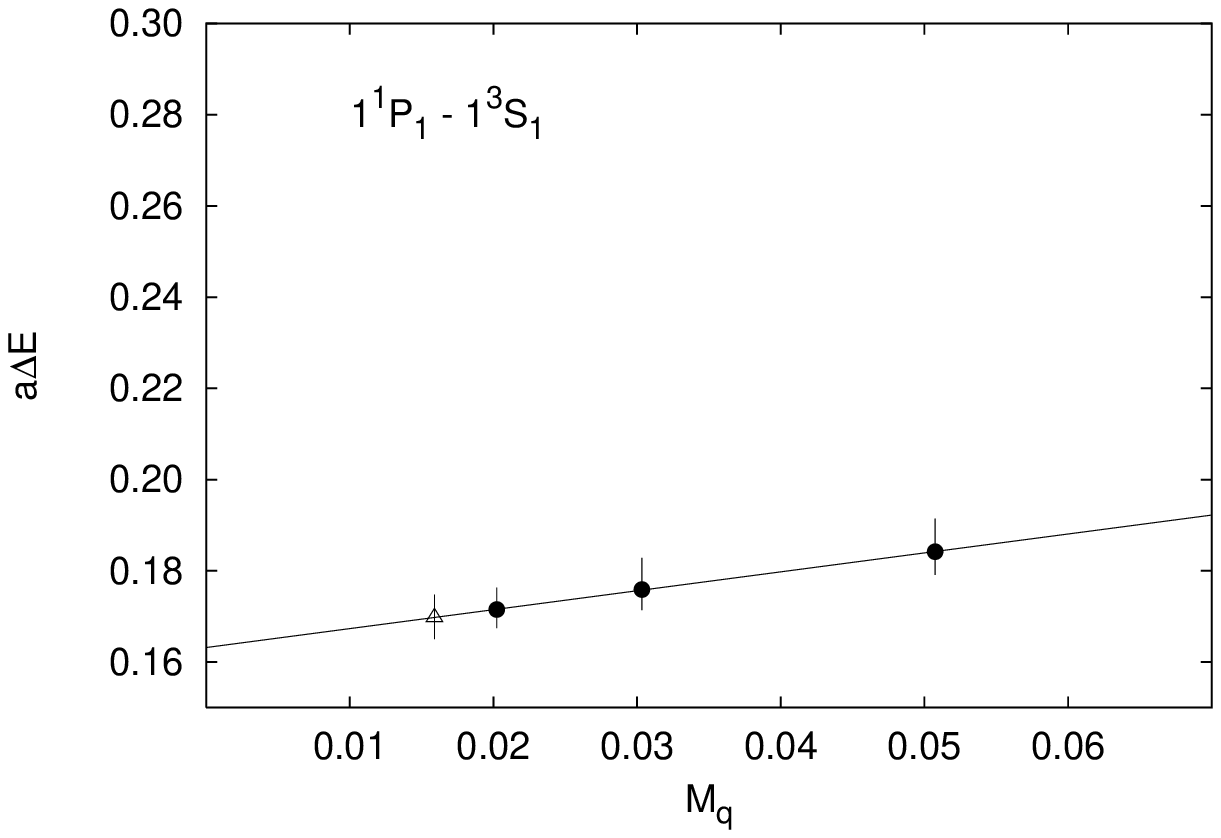,width=2.7in,height=2in}
{SESAM result for the $^1P_1-^3S_1$
splitting
as a function of the light quark mass.}  

Both  collaborations performed their unquenched calculations
with $n_f=2$, and extrapolated to $n_f=3$. Indeed, the NRQCD
collaboration 
found that fixing the lattice spacing using different
$\Upsilon$ splittings led to results which varied by $3~\sigma$ when
$n_f$ is taken to be zero. This discrepancy essentially 
disappears when $n_f=3$. So it
seems reasonable to expect that these new unquenched calculations
have the light quark effects well under control. We should point
out that neither collaboration took into account $SU(3)$ violation
, but these effects should be extremely small, given the size of
the bound state.

A larger error arises due to the uncertainty in the, yet to
be calculated, $n_f$ dependent piece in the relation \ref{relate}.
The SESAM group 
 varied the coefficient $C_2$ in \ref{relate}, between 1 and -1 
and found a $2\% -3\%$ variation in their result while the NRQCD
collaboration quote a $1.9\% $ error due to the uncertainty in $C_2$.
The SESAM collaboration quotes the results 
\begin{eqnarray}
&& \alpha^{(5)}_{\overline {MS}}(m_Z)=\nonumber \\ && \left\{ \begin{array}{l l}
0.1118~(10)(12)(5) & \mbox{${\bar \chi}-\Upsilon$ splitting} \\
  0.1124~(13)(12)(15) &\mbox{$\Upsilon^\prime
-\Upsilon$ splitting} \end{array} \right., 
\end{eqnarray}
where ${\bar \chi}$ is the spin average of the P wave states.
The first error is statistical while the last two are from the NRQCD
truncation
and uncertainty in the sea quark mass dependence.
The errors do not include any guess at the size of discretization
error stemming from the light quarks, which the authors estimate to
be at the $5\%$ level. Nor does this error include the uncertainties
due to our ignorance of the $n_f$ dependent piece of $C_2$, which
the authors estimate to be at the $2\%-3\%$ level.

The NRQCD collaboration quotes
\begin{eqnarray}
&& \alpha^{(5)}_{\overline{MS}}(m_Z)=\nonumber \\ && \left\{ \begin{array}{l l}
0.1174~(15)(19) & \mbox{${\bar \chi}-\Upsilon$ splitting} \\
  0.1173~(21)(18) &\mbox{$\Upsilon^\prime-\Upsilon$ splitting} 
\end{array} \right. 
\end{eqnarray}

The errors are due to lattice artifacts and perturbative truncation
errors, respectively. Notice that the two collaborations differ
by $3\sigma$. The SESAM collaboration attributes this to
light quark discretization errors. 

We see that these  calculations seem to
have the errors well under control. My only true concern, is the
issue of the convergence of the perturbative expansion at
these scales. Given the precision of the measurement, a calculation
of the complete $C_2$ coefficient would certainly increase the
confidence level.

\section{Onium Production} 
In \cite{BBL} it was pointed out that by combining 
perturbative factorization with NRQCD it is possible
to make ``rigorous'' \footnote{The calculation have a level
of rigor equivalent to those in semi-inclusive hadro-production where
one relies on perturbative factorization in the physical region.} 
predictions for onium production.
A general production process may be written as
\begin{equation}
d\sigma =\sum_n d\sigma_{i+j\rightarrow Q\bar{Q}[n]+X}\langle 0 \mid O^H_n
\mid 0 \rangle.
\end{equation}
Here $d\sigma_{i+j\rightarrow Q\bar{Q}[n]+X}$ is the short distance
cross section for a reaction  involving two partons $i$ and $j$, in
the initial state, 
and two heavy quarks, in a final state labeled by $n$, plus $X$. 
This part of the process is short distance dominated and completely
calculable in perturbation theory, modulo the possible structure
functions, in the initial state and may be considered as ``matching
coefficient''. 
The long distance part of the
process involved the hadronization of the heavy quarks in the
state $n$ into the hadron of choice $H$.
Indeed, the matrix element which is written as
\begin{eqnarray}
&&\langle 0 \mid O^H_n
\mid 0 \rangle= 
 \langle 0 \mid \psi^\dagger \Gamma^{n^\prime} \chi \mid\sum_X  H+X\rangle \nonumber
\\ & & \langle H+X \mid \chi^\dagger \Gamma^n \psi \mid 0\rangle.
\end{eqnarray}  
The tensor $\Gamma^n$ operates in color as well as spin space and also
contains possible derivatives. This tensor also 
 determines the order, in $v$ of the matrix
element. The size of the matrix element is fixed by determining
the perturbations necessary to give a non-vanishing result for
the time ordered product(selection rules). The matrix elements are
taken
between states of the effective theory, which are given by the
eigenstates of the dipole expanded version of \ref{Leff}.
Thus, at leading order there is no overlap between operators
with quantum numbers (usually labeled spectroscopically  $^{2S+1}L_J(1,8)$) 
which differ from those of the  state under consideration.
Which is to say if $H$ has the quantum numbers $^{2S+1}L_J$ then at
leading
order in $v$ color
singlet operators give
\begin{equation}
\langle 0 \mid O_{(1)}^H(^{2S^\prime
+1}L^\prime_{J^\prime})
\mid 0 \rangle \propto \delta_{SS^\prime} \delta_{L
L^\prime} \delta_{JJ^\prime},
\end{equation}
while all color octet operators give zero.  Higher order contributions
can be included by inserting higher multipole moment interactions into
the time ordered product.  For instance, the matrix element $\langle 0
\mid O^{J/\Psi}_{(8)}(^3 S_{1}) \mid 0 \rangle$, would scale as
$v^4$, since we need two $E1$ insertions at the the amplitude level,
each costing a factor of $v$. The first insertion neutralizes the
color, but it also changes $L$. So a second insertion is needed to
bring us back to the $S$ wave state. This is illustrated in figure 1.

\EPSFIGURE[h]{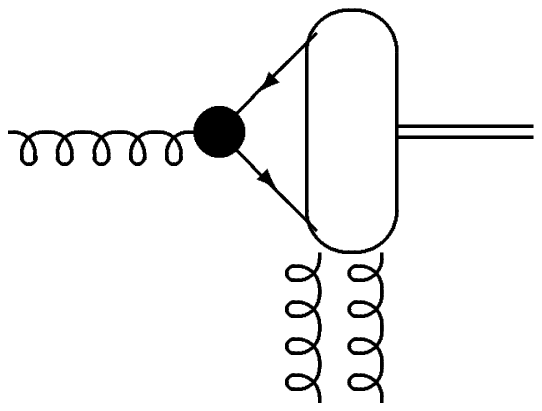,width=2.2in,height=1.8in}
{Gluon fragmentation into a $^3S_1$ state. The quark anti-quark pair
are created at a short distance scale in a relative octet state. 
They then propagate into
the $J/\psi$ state by emitting two soft $E1$ gluons.}  

For the color singlet matrix elements, we may use factorization, which
is typically correct up to $v^4$ corrections, \footnote{ This is true 
unless the $L$ of the state
created is different from the $L$ of the operator under consideration,
in which case the correction can be $O(v^2)$.} to write
\begin{equation}
\label{factor}
\langle 0 \mid O^H_n
\mid 0 \rangle=\langle 0 \mid \psi^\dagger \Gamma^n \chi \mid H
\rangle
\langle H \mid \chi^\dagger \Gamma^n \psi \mid 0 \rangle. 
\end{equation}
We may then interpret 
$\langle H \mid \chi^\dagger \Gamma^n \psi \mid 0 \rangle$ as
the ``wave function'' at the origin $\psi(0)$, or derivatives thereof,
in a potential model (with some factorization scale dependence). 
This parameter is the same parameter which
determines the annihilation rate and thus its value is easily
extracted from the data.

Note that if we were to drop all color octet contributions, then
we would end up with the old fashioned, color singlet model.
In this model the production rate is calculated by weighting the
partonic rate to produce two heavy quarks with zero relative 
velocity with the a potential wave function at the origin $\mid
\psi(0) \mid^2$. Here I wish to emphasize that this method of
calculation is indeed a {\it model}, in that there is no limit in
which it reproduces the full QCD calculation. On the other hand,
despite what might often be said, the NRQCD prediction is NOT
a model. In the limit where the quark mass is much larger than
the QCD scale the theory reproduces full QCD. Thus, those who call
the NRQCD prediction, the ``color octet'' model, have been either 
 gravely mislead
or are using a definition of the term {\it model} with which I am not
familiar.

The power of the NRQCD/factorization formalism for production lies in
the universality of the matrix elements $\langle 0 \mid O^H_n \mid 0
\rangle$. So that to make a prediction for a given process we must
first extract the appropriate matrix elements from another
process. This universality is a consequence of the factorization of
the short distance production process from the relatively long
distance hadronization process. At large $p_T$ production is
dominated by fragmentation processes \cite{BY}. In this case, the cross section
is written as the probability to create a nearly on-shell parton with
momentum $q$ such that $q_0 \gg \sqrt{q^2}$, times the probability of
the parton to fragment into the hadron of interest.  This latter
probability may be written in terms of an unphysical fragmentation
function $D^H_i(z)$ which gives the probability of forming a hadron
$H$ with momentum fraction $z$ from an initial parton labeled by $i$.
This factorization is analogous to case of factorization for
semi-inclusive light hadron production\cite{CSS}.  In the case of
onium production the fragmentation function itself factorizes into a
product of the probability for an energetic parton with invariant
mass on the order of $4m_Q^2$ to produce two heavy quarks in a state
$^{2S+1}L_{J}^{(1,8)}$ times the probability of this heavy quark state to
hadronize into the onium state of interest. Thus, as opposed to the
case of light hadrons, the fragmentation function can be written in
terms of a perturbative expansion in $\alpha_S(2m_Q)$ and an unknown
constant(s) (our matrix elements from above).  Writing the cross section
in this way makes it simple to resum large logs of the form
$\log{p_T/m_Q}$ (i.e. perform the Alterelli-Parisi running).  In this
case, the factorization holds up to corrections of order
$4m_Q^2/p_T^2$. 

For smaller values of $p_T$ fragmentation no longer dominates and one
must consider the fusion of partons into the final state. This
cross section is written as the probability to create a heavy
quark anti-quark pair, in a given configuration, times 
the hadronic matrix element.  In this regime higher
twist effects are expected to scale as powers of
$\Lambda/\sqrt{p_T^2+m_Q^2v^n}$ \footnote{ The first power correction
would most likely scale as $\Lambda^2$.}, though this has not been
fully explored to date ($n$ has not been fixed), since the quark pair
can be strongly influenced by the jet in the forward direction as
$p_t$ gets smaller. Thus,  our confidence in our calculation
dwindles as $p_T$ is reduced.  We
should thus keep in mind that when we test the formalism we are
testing more than just the convergence of the $\alpha_s$ and $v$
expansions. We are also testing factorization.
Thus, at least for production, and to a lesser degree for decays, 
we should really say that
we are testing the NRQCD/factorization formalism.

\subsection{The Universality of the Matrix Elements }

A first crucial test of the NRQCD/factorization formalism consists of
checking the universality of the matrix elements. There are many
places from which we may extract the relevant matrix elements.  The
first extractions were done using the Tevatron data.  Let us, for the
moment concentrate on the case of the $J/\psi$, where we presently
have most of our data.  As discussed above, at large $p_T$ gluon
fragmentation dominates and gives a leading order contribution
proportional to $\langle 0 \mid O^{J/\psi}_{8}(^3S_1)\mid 0 \rangle$,
as shown in figure 2, while at intermediate values of of $p_T$, the
rate becomes sensitive to $\langle 0 \mid O^{J/\psi}_{8}(^1S_0)\mid 0
\rangle$ and $\langle 0 \mid O^{J/\psi}_{8}(^3P_0)\mid 0 \rangle$.
Actually at lower values of $p_T$ the rate is sensitive to a
linear combination of these last two matrix elements. This combination
is usually defined as
\begin{equation}
M_k^{J/\psi}=\langle O^{J/\psi}_{8}(^1S_0)\rangle +{k
\over{m_Q^2}}\langle O^{J/\psi}_{8}(^3P_0)\rangle,
\end{equation} 
$k$ actually varies with $p_T$. The extraction of $M_k$ is complicated
by the fact that its value is very sensitive to the small $x$ gluon
distribution which makes it difficult to extract reliably. Varying
parton distribution function can change the extracted values by a
factor of four.  Therefore, I will concentrate on extractions of
$\langle O^{J/\psi}_{8}(^3S_1)\rangle$.
\vskip.1in
\TABULAR[ht]{|c |c |c |c|c|c |c |c|}{ \hline
$J/\psi$ &CTEQ2L&  CTEQ4L&CTEQ4M & GRVLO & GRVHO & MRS(R2) &MRSD0 \\ \hline
 \cite{CL}&-& -&-& -& -&-& $0.66 \pm0.21$ \\ \hline 
 \cite{BK}& -&$1.06\pm0.14^{+1.05}_{-0.59}$ &-&
$1.12\pm0.14^{+0.99}_{-0.56}$&-&$1.40\pm0.22^{+1.35}_{-0.79}$
& - \\ \hline 
 \cite{CS}& $0.33 \pm 0.05$ & -&-  & -& $0.34\pm0.04$&-&$0.21\pm0.05$ \\ \hline
 \cite{KK} & - & -& $0.27\pm0.04$& -&-&-&- \\ \hline
 \cite{SL}&$0.96\pm 0.15$&- &-&-&$0.92 \pm 0.11$& -&$0.68 \pm 0.16$ \\ \hline }{Extractions 
of $\langle 
O^{J/\psi}_{8}(^3S_1)\rangle$ in units of $10^{-2} ~GeV^3$. 
The first error is statistical, while
the second, when listed, is due scale dependence. 
}
\TABULAR[ht]{|c |c |c |c|c|c |c |c|}{ \hline
$\psi^\prime$ &CTEQ2L&  CTEQ4L&CTEQ4M & GRVLO & GRVHO & MRS(R2) &MRSD0 \\ \hline
 \cite{CL}&-& -&-& -& -&-& $0.46 \pm0.10$ \\ \hline 
 \cite{BK}& -&$0.44\pm0.08^{+0.43}_{-0.24}$ &-&
$0.46\pm0.08^{+0.41}_{-0.23}$&-&$0.56\pm0.11^{+0.54}_{-0.32}$
& - \\ \hline 
 \cite{CS}& $0.14 \pm 0.03$ & -&-  & -& $0.13\pm0.02$&-&$0.11\pm0.03$ \\ \hline
 
  }{Extractions 
of $\langle 
O^{\psi^\prime}_{8}(^3S_1)\rangle$ in units of $10^{-2} ~GeV^3$. 
The first error is statistical, while
the second, when listed, is due scale dependence. 
}

Several groups have extracted this matrix element
\cite{CL,BK,CS,KK,SL} for
the $J/\psi$ as well as the $\psi^\prime$.
I have  collected the various  extracted values from these references 
in tables  1 and 2. The extractions differed
in several respects. References \cite{CS,KK,SL} included  
a Monte Carlo  estimation
for initial state radiation, which we expect to become more
important for smaller values of $p_T$. 
Of these three references though, only \cite{SL} correctly accounted for
the Alterelli-Parisi evolution, which is seen to be important for
this particular matrix element. Noting the difference between
\cite{SL} and \cite{CS,KK}, we can see that the effect of resummation
is only important, as we would expect, for small $p_T$.
Only \cite{BK} made any estimates 
for the scale dependence of the results, which from the table
can be seen to be substantial. The extractions also differ in
the way in which they treat the interpolation between the
fragmentation
at high $p_T$ and the direct production at low $p_T$. 
I think that it would reasonable to say that the dominant source
of error will come from higher order corrections, as evidenced
by the large scale dependence found by \cite{BK}. 
It would be interesting to see the extent to which
 this scale dependence is reduced by including higher order effects.
However, given the present state of affairs, I don't think I would
be overly conservative if I were to say that there is a factor
of two uncertainty in the octet matrix element.

 At LEP, we might hope that we can get a better handle on the octet
matrix elements since the theoretical calculation is perhaps under
better control. In this case we need not worry about factorization
scale dependence in parton distribution functions or initial state
gluon radiation. However, there is a  complication do to
final state soft gluon emission. Indeed, $J/\psi$ production at LEP
is a beautiful example of  color coherence. Moreover, as opposed
to hadronic collision, soft gluon effects are completely
calculable in closed form in lepton initiated processes, as I will
know discuss. 

If one calculates the leading order differential cross section for
octet production \cite{Cho,BKLSI} one finds schematically 
\begin{equation}
{d\Gamma \over{dz}}(Z\rightarrow J/\psi+X) \propto \alpha_s^2
\log(M_Z^2/M_{J/\psi}^2)/z+~~~, 
\end{equation} 
where $z=2E_{J/\psi}/M_Z$ and the terms left off are less singular in
the
$z\rightarrow0$ limit. We see that the differential rate is
dominated by an infrared enhancement coming from the small $z$
region. This enhancement leads to large double logs in the total rate.
Given that $\alpha \log(M_Z^2/M_{J/\psi})\simeq 1$, a leading
order prediction necessitates a resummation of these logs. Such a
resummation can not be accomplished by standard ladder resummation
techniques, due to the color coherence of the soft gluon radiation
\cite{DDT}.  However, by imposing angular ordering it is possible to
write down an integral equation for the resummed fragmentation
function \cite{MUL}. The resummed differential rate calculated in
\cite{BLR} is shown in figure 3. Notice that the total rate is
dominated by the octet at small $z$. The peak at small $z$ is an
example of what is called the "hump-backed distribution" that arises
in calculations of jet-multiplicities \cite{DDT}. The prediction is
not valid at smaller values of $z$ due to the complete breakdown of the
perturbative expansion, as well the the breakdown of a saddle point
approximation used in \cite{BLR}.  Nonetheless, the region well 
below the peak 
contributes negligibly to the total rate. The data from LEP includes
feed-down from excited states, so that the rate is not proportional to
$\langle O_8^{J/\psi}\rangle$, but instead is proportional to a
combination of matrix elements. This is the drawback of this
study.  The rate is in fact proportional to the combination
\begin{eqnarray}  \langle \hat{O}_8^{\psi(m)}(^3S_1&&\!\!)\rangle \equiv 
\sum_{m}\langle O_8^{\psi(m)}(^3S_1)\rangle \times \nonumber \\
&&  BR(\psi(m)\rightarrow
J/\psi+X), 
\end{eqnarray} 
where $\psi(m)$ are excited states of the
$J/\psi$.  The authors of \cite{BLR} found 
\begin{equation} \langle
\hat{O_8^{J/\psi}(^3S_1)}\rangle=(0.019\pm0.005_{stat}\pm 0.010_{theory})
~GeV^3.  \end{equation}
The theoretical error quoted is
conservative, and includes the uncertainty due to relativistic
corrections, subleading logs and factorization scale dependence.  
If one compares
this with an analogous extraction from the Tevatron data, which has
errors on the order of $100\%$, one finds agreement.
Before moving on I should point out that the peak is almost purely
color octet. Thus the data seems to support the existence of this
channel with a matrix element of order that found at the Tevatron.

\EPSFIGURE[h]{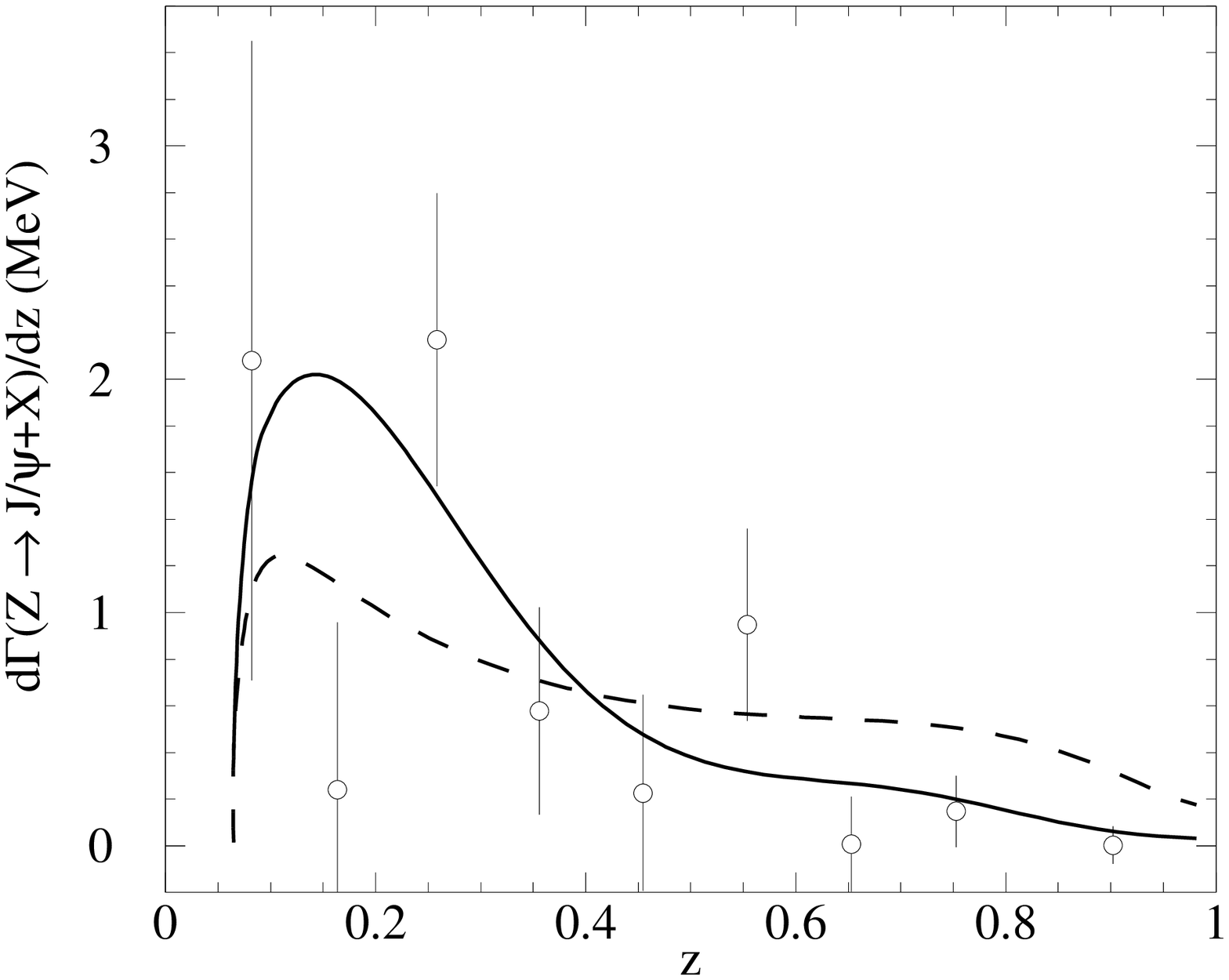,width=2.8in,height=2.2in}
{Results for $d\Gamma/dz$ at LEP taken from \cite{BLR}. The dashed
line shows the result without the resummation.}  

Before concluding our discussion of universality, I should address the
so-called "HERA-anomaly".  It has been pointed out that at HERA in the
large $\hat{z}$\footnote{$\hat{z} \equiv E_{J/\psi} / E_\gamma$ in the protons rest
frame}
 region, NRQCD predicts a sharp rise in the inelastic
$J/\psi$ direct photo-production cross-section \cite{CK,KLSI}.
This behavior is due to the fact that the $^1S_0$ and $^3P_J$
configurations
can be produced via a t-channel gluon at lowest order in $\alpha_s$.
A similar statement can be made of the $^3S_1$ configuration for
resolved photo-production. No such peak near $\hat{z}\approx 1$ is seen in
the
data leading to the aforementioned ``anomaly''.
However, as is now well appreciated, 
as $\hat{z}$ approaches 1, the theory for the spectrum becomes
intractable for several reasons.
Firstly, the endpoint region  is
sensitive to initial state 
intrinsic transverse momentum 
(equivalently initial state radiation\cite{KKII}). 
Indeed, the authors of 
 \cite{SMS} found
that by modeling the intrinsic transverse momentum with a Gaussian
distribution with $\langle k_T \rangle \approx 0.7 ~GeV$
they were able to fit the data. 
Secondly, as $\hat{z}$ approaches its endpoint ($\hat{z}>.75$)
the spectrum becomes 
sensitive to the long distance hadronization process. This
manifests
itself as a breakdown in the NRQCD expansion \cite{RW,MW}.
In \cite{BRW} it was shown that the rate may be written in the
following convoluted form
\begin{eqnarray}
&& \frac{d\sigma}{d\hat{z}} = \int\limits_{P^2_{T,min}}\!dP_T^2
\int\limits_0^1\!dx\int\!dy_+\,S(x,\hat{z},P_T^2)\,F_H[n](y_+)
\nonumber\\
&&\delta\!\left(s(1-\hat{z}) x- 
\frac{M^2 (1-\hat{z})+P_T^2}
{\hat{z}}  \right.\nonumber \\ 
&&\qquad\qquad\qquad \left. - 
\frac{M^2 (1-\hat{z})^2+P_T^2}{\hat{z} (1-\hat{z})}
y_+\right).
\end{eqnarray}
The lower cut on $p_T$ is necessary to eliminate the diffractive
contribution. The function $F_H[n](y_+)$ is a leading twist
distribution function defined in \cite{BRW} which physically accounts
for the momentum carried by the non-perturbative gluons. For any
non-zero $P_T$ we see that the expansion parameter close to
$\hat{z}=1$ is $y_+/(1-\hat{z})\sim v^2/(1-\hat{z})$. Thus, the NRQCD
expansion breaks down for $1-\hat{z} \sim O(v^2)$, because
higher-order terms in $v^2$ grow more and more rapidly as $\hat{z}\to
1$. Consequently, the NRQCD factorization approach makes no prediction
in the endpoint region and the discrepancy between leading order
predictions and  the data in this region does not allow us to
draw any conclusion on the relevance of color-octet contributions to
photo-production.  If one averages the $\hat{z}$-distribution over a
sufficiently large region containing the endpoint, the octet
mechanisms contribute significantly to this average. However, the
characteristic shape information is then lost and one has to deal with
the more difficult and uncertain question of whether the absolute
magnitude of the cross section requires the presence of octet
contributions, and whether their magnitude is consistent with other
production processes.
One might hope to learn more about the octet matrix element from
studying the smaller $\hat{z}$. However, there are large uncertainties
in the color singlet contribution \cite{BKV,K} which make the prospect
of learning much from this region rather bleak. 

\subsection{The question of polarization}

NRQCD predicts that at large transverse momentum, when gluon
fragmentation
dominates, the onium should be dominantly transversely polarized
\cite{CW}. This polarization arises as a consequence of the fact
that the gluon is nearly on shell and thus
is transversely polarized up to corrections of order $4m_c^2/p_T^2$.
This transverse polarization is inherited by the onium, since
the electric transitions preserve the spin of the heavy quarks.
There are corrections coming from hard gluons which can flip the
spin but these corrections turn out to be surprisingly small\cite{BRI}. There
are also spin symmetry breaking magnetic transitions which are
suppressed by $v^4$ compared to the leading order transition. 
Thus, we have the rather robust prediction that $\psi$ production
at large $p_T$ should be dominantly transversely polarized and this
polarization should increase with $p_T$. 

Presently there is data from CDF for both polarized $J/\psi$ and
$\psi^\prime$ production \cite{poldata}. The data on the $J/\psi$ shows
no signs of polarization. However, this data includes feed-down from
higher excited states, which we would expect would dilute the
polarization \footnote{ A quantitative study of this dilution would
be helpful.}. The $\psi^\prime$ sample is perhaps more surprising.
In this case we don't expect to have the dilution problem, and thus
if our approximations are good, we should see a rise in the fraction
of transversely polarized $\psi^\prime$'s as $p_T$ increases. 
The preliminary data does not seem to show this trend. Indeed, it
seems to indicate a trend towards {\it longitudinal} polarization 
with increasing $p_T$. Of course, the data still has large errors
and thus it is perhaps as little too early to jump to any conclusions.
However, the trend is rather disturbing.

\section {Conclusions and Outlook}

In this review, I have touched upon only two applications of
NRQCD. There are several other applications which were discussed
at this meeting \cite{beneke}. As far as using NRQCD for extracting
$\alpha_s$, I think that the next important step will be to
try to get a better handle on the perturbative relation between
$\alpha_{\overline{MS}}$ and $\alpha_P$, which will be a rather
Herculean task. However, given that we believe that the series is
asymptotic, there will always be doubts as to how well the series
is behaving, and given the accuracy of these calculations, one 
always worries. 

As far as the NRQCD/factorization formalism, as applied to the charmed
system, is concerned we are presently in a state of
uncertainty\footnote{I have not discussed the $\Upsilon$ system due to
space limitations and the fact that until recently the data was rather
scant.}. Let us review the successes and failures of NRQCD in this
context.  To begin with let's consider predictions for cross-sections.
The data at the Tevatron for $\psi^\prime$ seems to necessitate a
color octet contribution \cite{BF}. The spectrum is well fit, once the
octet contributions are included.  This is not a terrible surprise
since there are two octet contributions one which behaves as $1/p_T^4$
the other as $1/p_T^6$, nonetheless this is encouraging.  
The overall normalization of the
prediction is uncertain at about a factor of two level.
At LEP the
theory is under better control, but the data is much more sparse. The
theory predicts a large hump at small energies due to the octet
contribution which seems to be supported by the data. A combined LEP
analysis of the spectrum would certainly be welcomed. As far as the
HERA data is concerned the data can be well fit without the
contribution from the octet channel at next to leading
order\cite{K,KZSZ}. The expected rise at large $\hat{z}$ from the
octet, which is not seen, should not trouble us at this time, for
reasons discussed above. Octet contributions to production in fixed
target experiments\cite{BRII,GS,TV}, as well as in $B$ decays
\cite{KLSII,BRIII} have been analyzed. These studies also seem to
yield matrix elements of order $10^{-3}~GeV^3$.  However, we would
expect higher twist contributions in these reactions to be especially
important.

As far as the qualitative size of the extracted matrix element is
concerned, NRQCD seems to have it right. If we use the relation
\ref{factor} to extract the singlet production matrix element from a
decay process, we find that the ratio of octet to singlet matrix
elements is of order $10^{-3}$. Given that we expect $v^2 \simeq 0.3$
in the $J/\psi$ system and that according to NRQCD scaling rules the
ratio of these matrix elements should scale as $v^4$, we seem to have
qualitative agreement. However, I would like to point out that the
relation \ref{factor} itself has never been tested. Which is to say,
we have yet to extract the size of the singlet production matrix
element.  Indeed, all extractions of octet matrix elements assume the
relation \ref{factor} and then use the decay matrix element, either
from a potential model, or from an NRQCD extraction. One could worry
that the production singlet matrix element is actually smaller than
expected. As far as I can tell, this can not be ruled out by the
data, at least for the $J/\psi$ \footnote{Here I am not including the
data from fixed targets, since the higher twist effects could be
large.  Also, at HERA for $\hat{z}\approx 0.5$, it might be
difficult, but given the uncertainties, not impossible.}. One easy way
to check, would be to extract the singlet production matrix element
at CLEO, where the singlet dominates, away from the endpoint
\cite{BS}. While I don't expect the singlet to be anomalously small,
it is still important to extract the singlet production matrix
element, since the perturbative series for decays seems to be very
poorly behaved \cite{BSS}.

For more quantitative results, we should stick to the large $p_T$
processes where we feel more confident that higher twist effects are
small. This essentially leaves us with the Tevatron data and the LEP
data. The Tevatron extraction is plagued by large factorization scale
dependence, which could indicate a poorly behaved perturbative
series. While the next to leading order correction to the total
cross-section has been calculated \cite{PCGMM}, a next to leading
order calculation of the differential cross-section $d\sigma/dp_T$
would allow us to study higher order corrections without worrying
about higher twist contamination.  The LEP extraction, on the other
hand, is limited by statistics. Thus, at least for now, testing the
universality of the matrix elements, at an accuracy level beyond
$100\%$ is difficult.

Finally, there is the issue of polarization. In addition to the
studies at the Tevatron, polarization has also been investigated in
fixed target experiments, where the theory is compatible with the
experiment within errors, given the large uncertainty in the octet
matrix elements. There have also been interesting theoretical studies
of $J/\psi$ polarization in $B$ decays \cite{FHMN}, LEP \cite{BKLSII}
and at HERA for photo-production \cite{BKV} as well a lepto-production
\cite{MF}, which could in the future provide useful
information. Presently, we should be concerned with the trend in the
Tevatron data.  If, once the statistics improves, the data continues
to show no trend towards transverse polarization at larger $p_T$ then
we have a very intriguing puzzle on our hands. Where could the theory
be going wrong? A naive guess would be the spin symmetry is badly
violated. Which is to say that perhaps the magnetic transition
operators are not as suppressed at we think. However, while possible,
this seems unlikely given the fact that it seems to work so well in
the $D$ meson system.  If spin symmetry were badly violated, then we
would expect the splitting between the $D$ and the $D^*$ to be much
larger than it is. Of course, the matrix element which determines this
splitting is different than those which induce the spin flipping
transitions in $\psi^\prime$ production. So we can't draw any hard
conclusions from smallness of the splitting. The next naive guess
would be that there are large perturbative corrections to the
fragmentation which lead flip the spin. But, as discussed above, these
corrections have been calculated \cite{BRII}, and are indeed small.
Another possibility is that gluon fragmentation does not dominate at
large $p_T$. For instance, it could be that the production singlet
matrix elements $\langle O^{\psi^\prime}_{1}(^3S_1)\rangle$ is 30 times
larger than expected from relation \ref{factor}. The problem with this
proposal is that it would mean that the LEP data is too small by an
order of magnitude at large energies. Thus, there is no obvious way to
depolarize the $\psi^\prime$ at large $p_T$, no less get them to be
longitudinally polarized. I should stress, that the prediction for
polarization is based on very basic assumptions, which have been
tested in other context many times. Those assumptions are:
Factorization at large $p_T$, spin symmetry, and the standard parton
model assumptions which are routinely tested in Drell-Yan
processes. These assumptions are ``derivable'' from QCD, and have been
tested repeatedly.  So if the data persists we have a real puzzle on
our hands. Much more light will be shed on this problem as more data
streams in. With more upsilons  we may hope to gain information as
well, since in that system we believe all of our expansions will be
better behaved. However, we will have to go to larger values of $p_T$
to see the expected transverse polarization\cite{T}.  Nonetheless we still
should still see the trend.

\acknowledgments

I'd like to thank Colin Morningstar, Junko Shigemitsu and Achim Spitz
for discussions.  I would also like to thank Achim Spitz for supplying
me with the SESAM plots. I wish to also thank my collaborators on this
subject: Martin Beneke, Glenn Boyd, Ben Grinstein, Adam Leibovich,
Mike Luke, Aneesh Manohar and Mark Wise.  Finally, I'd like to thank
the conference organizers for running a worthwhile and enjoyable
conference.

\end{document}